\documentclass[11pt]{article}
\usepackage{epsfig}
\usepackage{amsmath}

\title{Statistical properties of one dimensional ``turbulence''.}

\author{Michel Peyrard and Isabelle Daumont\\
Laboratoire de Physique, Ecole Normale
Sup\'erieure de Lyon, \\
46 allée d'Italie, 69364 Lyon Cedex 07, France.}

\begin{document}

\maketitle

\begin{abstract}
We study a one-dimensional discrete analog of the von K\'arm\'an flow,
widely investigated in turbulence. A lattice of anharmonic oscillators
is excited by both ends in order to create a large scale structure in
a highly nonlinear medium, in the presence of a dissipative term
similar to the viscous term in a fluid. This system shows a striking
similarity with a turbulent flow both at local and global scales. 
The properties of the nonlinear excitations of the
lattice provide a partial understanding of this behavior.

\end{abstract}

\section{Introduction}
\label{sec:intro}

Although studies of turbulence have been carried out for decades, they
go on now-days because the problem is very hard and challenging but new
ideas have emerged in the last few years from the investigations of
statistical properties of turbulence, both at a local or a global
scale. First, studies of {\em local} longitudinal velocity differences
measured at distance $r$ showed that their probability densities are
very well described by a theoretical model based on nonextensive
statistical mechanics \cite{BECKLEWIS}, which, in the context of
turbulence, can be justified by a phenomenological model based on
dynamical foundations \cite{BECKPHYSA}. Second, interesting properties
showed up in the analysis of the statistical properties of {\em
global} quantities such as the power consumption of the confined
turbulent Von K\'arm\'an flow \cite{Pinton} and they culminated with
the observation of the universality of rare fluctuations in turbulence
and critical phenomena \cite{Bramwell}. 

These recent developments are based on an analysis of experimental
results \cite{BECKLEWIS,Pinton}. Indeed only such data contain the
full features of turbulence, but for an understanding of the results
it would be desirable to have a simple model, showing the basic
properties of turbulence, that would be amenable to very detailed
studies.  The search for simple models of turbulence lead to the
introduction of the so-called shell models \cite{ShellM} which
describe the energy exchange mechanism between various scales. However
shell models are not suitable to study properties that depend on space
or to analyze how the dynamics can lead to the observed statistics.

\smallskip
This is why it might be interesting to introduce models of turbulence
that are sufficiently simple to allow a detailed study, and
nevertheless preserve a spatial structure.  In this letter we
introduce such a model derived from studies on one-dimensional
nonlinear lattices. These systems show various features which are also
found in turbulence, and, in particular an exchange of energy between
various spatial scales which tends to cause a flow of energy from
large scales (long wavelengths) to smaller scales, through the
mechanism of modulational instability \cite{Kivshar}.  The idea to use
one-dimensional systems as simple models for turbulence is not new and
continuous one-dimensional media have been introduced to study wave
turbulence \cite{DYACHENKO}, but these studies only lead to weak
turbulence. Discrete lattice are attractive for two reasons. First
they are simpler since they contain only a limited number of degrees
of freedom (and therefore modes), but, more importantly because they
have exact localized solutions which could play the role of vortices,
the discrete breathers \cite{MACKAY94}, opening the possibility to
study an analogous of fully developed turbulence.

\smallskip
Many interesting experimental results on turbulence have been obtained
with a simple model flow, the Von K\'arm\'an flow produced in the gap
between coaxial disks \cite{Zandberg}.
We investigate here a nonlinear one-dimensional lattice excited by
both ends, in a configuration which drives a large scale structure in
the system, similar to the vortex of the Von K\'arm\'an flow and we
show that it exhibits remarkable similarities with turbulence, both
for local and global properties.

\section{The one-dimensional lattice model.}
\label{sec:model}

The system that we study is a one-dimensional chain of anharmonic
oscillators, coupled to each other by harmonic interactions. We have
chosen Morse oscillators because, the presence of a cubic term in its
expansion leads to a suitable nonlinear coupling between the modes,
and moreover the lattice of
Morse oscillators has localized oscillatory solutions which have
interesting properties \cite{NLloc}.
In analogy with the Von K\'arm\'an flow, we want to drive
a finite lattice from both ends, i.e. to inject energy in the system
through an external driving. As in actual turbulence experiment where
the viscosity of the fluids contributes do dissipate energy, we also
need to introduce a dissipation mechanism so that the system can reach
a steady state. In analogy with the dissipation term $\nu \nabla^2 v$
of the Navier Stokes equations, where $\nu$ is the kinematic viscosity,
we introduce a dissipation term proportional to the second order
finite difference of the velocity of the oscillators, so that the
equations of motion of the nonlinear lattice are
\begin{equation}
\label{eq:motion}
\ddot y_n - K (y_{n+1} + y_{n-1} - 2 y_n) - 2 \omega_0^2 e^{- y_n}
\left( e^{- y_n} - 1 \right) - \nu (\dot y_{n+1}
+ \dot  y_{n-1} - 2 \dot y_n) = 0 \; ,
\end{equation}
where $y_n$ denotes the position of the $n^{\rm th}$ oscillator, $K$
is the coupling constant between neighboring sites, $\omega_0^2$
determines the depth of the Morse on-site potential, and
 $\nu$ is a coefficient which plays the role of the kinematic
viscosity. As usual, the dot designates a time derivative.
 
Similarly to the Von
K\'arm\'an flow where the medium is excited from both ends by a set
of rotating disks which create a large scale structure in the system,
we consider a finite lattice of $N+2$ particles and we excite it by
imposing the displacements of the two end particles, particle $0$ and
particle $N+1$. Their motion is chosen according to a linear mode
of the lattice, with an amplitude $A$,
a wavevector $\kappa =  p \pi / (N+3)$, where $ 1 < p < N+1$ is an
integer, and a frequency $\omega = \sqrt{2 \omega_0^2 + 4 K \sin \kappa / 2}$
(assuming that particles $-1$ and $N+2$ are fixed).
Therefore the
driving of the system is defined by two parameters, $p$ which fixes
its wavevector and frequency and $A$ which determines its
amplitude. All simulations start from a lattice at rest and the
equations of motion are integrated with a fourth-order Runge-Kutta
scheme. 

\begin{figure}
\begin{center}
\mbox{\epsfig{file=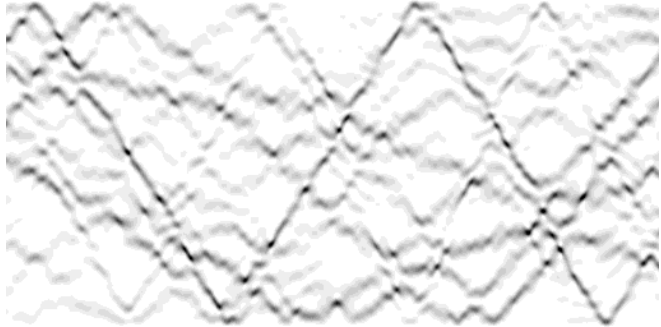,angle=90}}
\end{center}
\caption{
Energy density in
a $N=256$-particle lattice excited by a single mode
with $p=8$, $A=0.12$.
The parameters of the model are $\omega_0^2 = 1$, $K = 1$,
$\nu = 5~10^{-4}$.
The spatial dependence of the energy per unit celle $e_n(t)$ is shown with a
grey scale (white corresponding to 0 and black to a high energy
density). The horizontal axis corresponds to time and extends
over a time interval of 20 480 t.u., and the vertical
axis gives the position $n$ in the lattice.
}
\label{fig:grey}
\end{figure}

Indeed, for such a nonlinear lattice, the fact that the {\em driving}
corresponds to a single mode of the system does not imply that its
{\em response} will be merely a single mode. Similarly to the Von
K\'arm\'an flow, driven by two co-rotating disks, which has a very
complex response, the energy density nonlinear lattice also shows a
complex spatio-temporal pattern,
as displayed in Fig.~\ref{fig:grey}. 
Localized modes that move in the system, interact
with each other, and die out due to dissipation are observed, and they
appear to play a role very similar to the vortex filaments in fluids.
The point of this study is to observe and analyze the response of the
nonlinear lattice to a simple excitation. More complex cases could
easily be imagined, such as an excitation with two different
frequencies at both ends, which would be similar to some fluid
experiments where the two disks have different velocities, but, as
shown below the one-frequency driving is sufficient to yield
interesting and puzzling results.

\section{Local analysis.}
\label{sec:local}

Numerical investigations of spatial the Fourier spectrum of the energy
density $e(\kappa,t)$ in this simple model have exhibited properties
analogous to turbulence, such as a  power law spectrum 
$e(\kappa) = E_0 \kappa^{\alpha}$ with $\alpha \approx 2.2$
which is remarkably
insensitive to the parameters of the model or of the excitation in a
very broad range of parameters \cite{DAUMONTPEYRARD}.  
When the excitation amplitude
increases even more, a transition to a spectrum decaying exponentially
versus $\kappa$ is found when the large scale structure created by the
excitation breaks down into many smaller scale localized modes, as
shown in Fig.~\ref{fig:grey}. This is the regime that we study
in this letter. In
actual turbulence experiments the energy density cannot be directly
measured, but the local velocity is easily accessible. A map of the
velocity field would require an array of probes that perturb the flow,
and this is why the investigations only evaluate velocity
differences measured in 2 points separated by a distance $r$, which
provide some information on the spatial structure of the flow, and are
also related to the viscous dissipation. For the nonlinear lattice it
is easy to calculate an equivalent quantity in the numerical
simulations $
u(\delta n, t) = \dot y_{n_0 + \delta n} (t) - y_{n_0} (t)
$,
where $n_0$ is a fixed reference site and $\delta n$ is an index that
measures the distance between the two points, in units of the lattice
spacing.
It should be stressed that this quantity is local because there is no
spatial averaging over the reference site position $n_0$.

In a typical numerical experiment, we simulate the dynamics of the
system for a time interval of $2~10^7$ time units (t.u.). The first
$5~10^5$ t.u.\ are discarded in order to allow the system to reach a
steady state before starting the recording of $u(\delta n, t)$. Then
we store $u(\delta n, t_i)$ at times $t_i$ separated by an interval
$t_{i+1} - t_i = 40$~t.u.\ which is large enough to ensure that
successive values are statistically independent. This gives us a set
of 375000 values $u_i$ for each $\delta n$. We evaluate the reduced
probability density of this distribution by first calculating $\tilde
u_i = (u_i - \langle u_i \rangle)/ \sigma$ where $\langle \cdot
\rangle$ designates a time average and $\sigma = \sqrt{ \langle u_i^2 \rangle
- \langle u_i \rangle^2} $ is the variance of the distribution. Then
we make  a normalized histogram of the 
$\tilde u_i$ values in 100 bins to get the normalized probability
density $\tilde P (\tilde u)$.

\begin{figure}
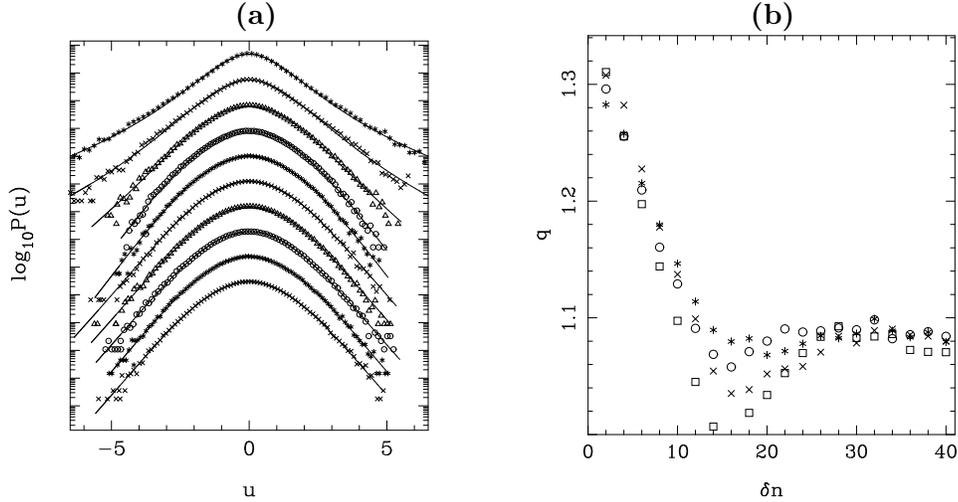

\begin{tabular}{cc}
\hspace{1.0cm}\textbf{(a)} & \hspace{1.7cm}\textbf{(b)} \\
\mbox{\epsfig{file=htsallis302.eps,width=5.6cm}} &
\mbox{\hspace{1.0cm}\epsfig{file=qtsallis301234.eps,width=5.6cm}} \\
\end{tabular}
\caption{(a) Probability density functions of the velocity differences
for $\delta n = 2$ (top plot) to $\delta n = 40$ (bottom plot), by
steps of $\delta n = 4$. The points correspond to the histograms
obtained by numerical simulations and the full lines are the fits of
$\tilde P(\tilde u)$ according to Eq.~(\protect\ref{eq:pu}).
The parameters of the model are $N=256$, $\omega_0^2 = 1$, $K=1$,
$\nu = 10^{-4}$ and $A = 0.12$. The reference site is $n_0 = 57$.
For better lisibility, each distribution is shifter by -0.5 units.
(b) Variations of the fitting parameter $q$ versus $\delta n / \nu$
for different viscosity parameters ($\nu = 5~10^{-5}$, circles
$\nu = 1~10^{-4}$, stars, $\nu = 5~10^{-4}$, x, $\nu = 2~10^{-3}$, squares)}
\label{fig:tsallisfit}
\end{figure}

Figure \ref{fig:tsallisfit}a shows such histograms for $\delta n$
varying from $2$ to $40$. The analogy with similar figures deduced
form experiments in fluid turbulence is striking (see for instance
Fig.~1 in Ref.~\cite{BECKLEWIS}). The same scenario from almost
Gaussian distribution at large distances to stretched distributions at
small distances is found. The analogy can be made more quantitative by
analyzing the shape of the histograms. The curves of
Fig.~\ref{fig:tsallisfit}a are very well fitted by the probability
distribution deduced from nonextensive statistical mechanics
\cite{BECKPHYSA,BECKPRL} 
\begin{equation}
\label{eq:pu}
\tilde P (\tilde u) = \frac{1}{Z_q} \frac{1}
{ \left[ 1 + \frac{1}{2} \beta (q - 1) {\tilde u}^2
\right]^{\frac{1}{q-1}}}
\; ,
\end{equation}
with $\beta = 2 / (5 - 3 q)$, determined by the condition that the
reduced probability distribution should have variance 1.
As $Z_q$ is a normalization constant, it should be noticed that each
curve is fitted with {\em only one} free parameter, the parameter $q
\ge 1$,
which, in the formalism of nonextensive statistical mechanics measures
the degree of nonextensivity. With $q \to 1$ ordinary statistical
mechanics is recovered and the probability distribution tends to a
Gaussian. In the case of turbulence, a formula that generalizes
(\ref{eq:pu}) has been derived to take into account the slight
skewness of the distributions \cite{BECKPHYSA}. The distributions
obtained for the nonlinear lattice also have a slight skewness so that
the extended formula improves the fit. However, as it is already very
good with Eq.~(\ref{eq:pu}) we sticked to the expression that
minimizes the number of parameters.

The variation of $q$ versus $\delta n$ is shown in
Fig.~\ref{fig:tsallisfit}b for two values of the viscosity $\nu$ and
a mean amplitude $\langle y \rangle \approx 1.1~10^{-2}$ in both
cases. As for the shape of the probability densities, the similarity
with the experimental results on turbulence is strong. At small
distances $q$ is significantly larger than 1 ($q \approx 1.3$ in our
case) and it decays toward a value close to 1, but above 1, at large
distances ($q \approx 1.08$ in our case). In turbulence experiments
the variation of $q$ versus distance is essentially determined by the
rescaled distance $r/\eta$, where $\eta$ is the Kolmogorov scale which
is the typical size of the smallest vortices. In the discrete
nonlinear lattice, the size of the smallest breathers is of the order
of the lattice spacing, and, in the low viscosity range (high
``Reynolds number'') that we are investigating here, it does not
vary significantly with the viscosity. This is why $q$ only depends on
$\delta n$ for the lattice as shown in Fig.~\ref{fig:tsallisfit}b.

\section{Global properties of the dynamics.}
\label{sec:global}

Our aim in this section is to examine to what extend the
one-dimensional nonlinear lattice that we have investigated also shows
the same ``universal'' statistical properties that where observed in
critical magnetic systems, self-organized criticality or turbulent
flows when one investigate global quantities.
\begin{figure}
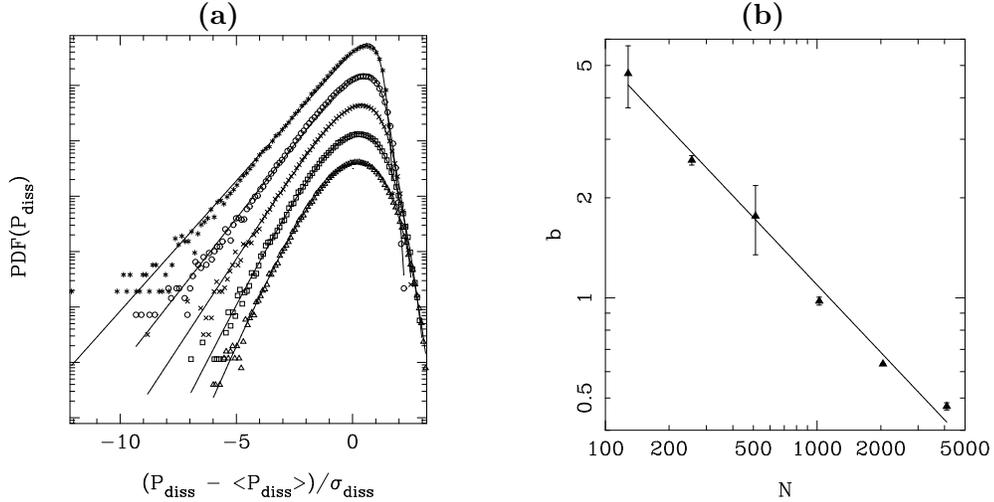

\begin{tabular}{cc}
\textbf{(a)} & \hspace{1cm}\textbf{(b)} \\
\mbox{\epsfig{file=multih.eps,width=5.6cm}} & \hspace{1cm}
\mbox{\epsfig{file=logbversuslogn.eps,width=5.90cm}} \\
\end{tabular}
\caption{ a) Probability distribution function (PDF) of the dissipated
power $P_{\mathrm diss}$ for different system sizes: from top
to bottom $N=256, 512, 1024, 2048, 4096$ The parameters of the
model are $\omega_0^2 = 1$, $K = 1$,
$\nu = 5~10^{-5}$. The points correspond to histograms deduced form the
numerical simulations and the full lines are fits with the generalized
Gumbell distribution. For a better lisibility, each histogram is
shifted by -0.5 unit along the vertical axis.
b) Variation of the parameter $b$ of the generalized Gumbell
distribution versus the system size $N$ in a log-log scale. The error
bars correspond to extreme values obtained in different simulations.  }
\label{fig:pdf}
\end{figure}

In actual turbulence experiments, a global quantity which can be
easily evaluated is the power injected by the engines driving the
flow, for instance through the disks in the Von K\'arm\'an experiment.
In the numerical simulations we have however the freedom to evaluate
other global quantities directly in the bulk of the system. The power
dissipated in the bulk
\begin{equation}
\label{eq:pdiss}
P_{\mathrm diss} = \sum_{n=N/4 + 1}^{n=3N/4} F_v(n) \, \dot y_n
= \nu (\dot y_{n+1} + \dot y_{n-1} - 2 \dot y_n ) \, \dot y_n
\end{equation}
(where $F_v$ is the viscous force) is particularly interesting because
it probes the dissipation, an essential physical process for
turbulence, at a global scale without being affected by the boundary
effects. To avoid any perturbation due to boundaries, 
we evaluate $P_{\mathrm diss}$ by
summing all the dissipative contribution from $n=N/4 +1$ to $3N/4$,
i.e. in the central half of the system. 

As $P_{\mathrm diss}$ is a global quantity, one
could think that it would be smooth and featureless due to averaging.
This is not the case. In
addition to small amplitude fluctuations, the signal shows sharp
peaks corresponding to strong bursts of dissipation. This feature
shows up on the normalized probability distribution function (PDF)
obtained from an histogram of the recorded values of $P_{\mathrm
diss}$ in numerical simulations. Figure \ref{fig:pdf} show such 
histograms 
plotted in logarithmic scales as a function of $\theta = ( P_{\mathrm
diss} - \langle P_{\mathrm diss} \rangle )/ \sigma$ where
$\langle P_{\mathrm diss} \rangle$ is the time average of the
dissipated power and $\sigma_{\mathrm diss}$ its standard deviation.
The dissipation bursts give rise to a highly asymmetric PDF with a
large tail on the negative side. The dissipation bursts are
associated to collisions of the localized excitations, the
discrete breathers, that are
visible on Fig.~\ref{fig:grey}.
Using an approximate expression of the breathers 
we can evaluate, in the continuum limit, the
time average of the power dissipated by a breather.
It shows that the dissipation grows with the third power of the
breather amplitude, so that the large
peaks observed during breather collisions generate strong
bursts of dissipation.

Figure \ref{fig:grey} showing the space-time evolution of the energy
density illustrates a typical situation where one observes
localized structures having a long life-time and a
complex pattern of interactions. They travel on long distances
introducing long range correlations in the dynamics of
the lattice. In the context of the general studies of correlated
systems which have been performed recently \cite{Bramwell}, systems
showing such correlations have been called ``inertial systems''  and
they display a strongly non Gaussian probability distribution function
of some global quantities. As shown by Fig.~\ref{fig:pdf}, this is
also true for the nonlinear lattice that we investigate here.
The strong asymmetry of the PDFs is
similar to the observations made on
turbulence and critical phenomena \cite{Pinton,Bramwell}. In order to
evaluate this quantitatively, we have fitted the numerical PDFs with a
generalized Gumbell distribution
\begin{equation}
\label{eq:gumbell}
\Pi (\theta) = w \exp \left[ a \left( b \theta - e^{b \theta} \right)
\right] \; ,
\end{equation}
where $w$ is the normalization factor and $a$, $b$ are parameters of
the fit.
Figure~\ref{fig:pdf} shows that this expression gives a very
good fit of the observed distributions, and is able to describe
properly their evolution as a function of the system size $N$.
In the limit
$ b\to 0$ the distribution (\ref{eq:gumbell}) tends to a Gaussian
distribution $\Pi (\theta) = w' \exp [ - \theta^2 / (2 \sigma_0^2) ]$
with a standard deviation $\sigma_0 = 1/(a b^2)$.
The value $b=1$ corresponds to the standard Gumbell
distribution observed in a turbulent flow \cite{Pinton}, and larger
values of $b$ correspond to even larger deviations from Gaussianity.
Figure \ref{fig:pdf} shows that, even for large systems
($N=4096$) the fluctuations of $P_{\mathrm diss}$ stay highly non Gaussian.

The origin of this non-Gaussian behavior can be understood from the
same argument as for the case of turbulence or critical phenomena
\cite{Bramwell}: there are long range correlations in the system which
cannot be divided into statistically independent regions. However the
origin of these long range correlations appears to be different. While
in the vicinity of a phase transition correlations are due to the
existence of structures at all scales, in the present case, we observe
{\em small scale} structures (localized modes) which can move on long
distances and preserve their identities in collisions (see
Fig.~\ref{fig:grey}). However the localized modes are {\em not}
solitons and are therefore perturbed by the collisions, which limits
the correlations to a finite range. As a result the PDF of the
dissipated power depends on the size of the system and on the density
of localized modes, both of which determine the average number of
collisions that a localized excitation experiences when it travels
inside the part of the lattice in which we compute the dissipated
power (i.e. along $N/2$ sites).

The effect of the size of the system is shown in
Fig.~\ref{fig:pdf}b which shows that $b$ decreases with $N$
approximately with a power law $b \propto N^{-0.67}$ for $128 \le N
\le 4096$ so that the approach of a Gaussian distribution for large
systems is very slow. The study of the effect of the mean amplitude
of the displacement, which is correlated to the density of localized
modes in the system, shows that a strong
non-Gaussianity of the power dissipated in the system persists
for variations of $\langle y \rangle$
that extend on more than two orders of magnitude.

Consequently, contrary to the case of turbulence where a universal form
of the fluctuations of a global quantity, independent of the Reynolds
number, was found \cite{Pinton}, the shape of the PDF of the
power dissipated in the nonlinear lattice changes with the size of the
system or amplitude of the excitation. However it stays strongly
non-Gaussian and well described by a generalized Gumbell distribution
for all the cases that we could investigate. One may wonder what are
the differences between the turbulent flow and our model system. One
possibility is the existence in the turbulent flow of structures at
all scales, including the integral scale, i.e.\ flow patterns that
have a characteristic size equal to the size of the system, which is
not true for the nonlinear lattice. One may ask however whether larger
Reynolds number would not break the large vortex into smaller
filaments, leading to a gradual change in the shape of the
PDF. Such a regime, if it exists, is probably beyond experimental
possibilities, while, for the one-dimensional nonlinear lattice, the
simulation can examine a very broad range of dynamical behaviors.

\section{Conclusion}
\label{sec:conclusion}

In this work we examined a simple model related to
turbulence. The statistics of its local and global properties are
remarkably similar to the experimental results of actual turbulence so
that it could be useful for its understanding because it is
sufficiently simple to allow detailed studies and test ideas.
The remarkable features of the local and global properties
could perhaps be understood in a single framework.  
Global properties of turbulence have been analyzed by
Bramwell et al. \cite{Bramwell} by  pointing out that the flow cannot
be divided into mesoscopic regions which are statistically
independent. Such a ``long range interaction'' is precisely what leads
to non-extensive thermodynamics \cite{TSALLIS} which models so well
the local properties.

Acknowledgements.

Part of this work has been supported by the EU Contract
No. HPRN-CT-1999-00163 (LOCNET network). M.P. wants to thank
T. Dauxois and B. Castaing (ENS Lyon) for helpful discussions and
E.D.G. Cohen (Rockefeller University) 
for an illuminating lecture that motivated part of this
work.

\end{document}